\documentclass[pra,twocolumn,nofootinbib,floatfix]{revtex4-1} 
\usepackage[usenames]{color}
\usepackage{hyperref}

\usepackage{bbold}

\usepackage{hhline}

\usepackage[utf8]{inputenc} 
\usepackage[T1]{fontenc}
\usepackage[pdftex]{graphicx}

\usepackage{amsmath, amsxtra, amssymb, mathtools, isomath, xspace, textcomp} 

\usepackage[british]{babel}

\usepackage{booktabs} 
\usepackage{multirow}

\newcommand{\ket}[1]{\ensuremath{\left| #1 \right\rangle}} 
\newcommand{\bra}[1]{\ensuremath{\left\langle #1 \right|}} 
 \newcommand{\down}{\ensuremath{\ket{\downarrow}\ }} \newcommand{\up}{\ensuremath{\ket{\uparrow}\ }}
 
   \usepackage{soul}

\long\def\symbolfootnote[#1]#2{\begingroup%
  \def\thefootnote{\fnsymbol{footnote}}\footnotetext[#1]{#2}\endgroup}

\begin{document}

\title{Microscopic observation of magnon bound states and their dynamics}

\author{Takeshi Fukuhara$^{1,*}$}%
\author{Peter Schau{\ss}$^{1}$}%
\author{Manuel Endres$^{1}$}%
\author{Sebastian Hild$^{1}$}%
\author{Marc Cheneau$^{1,2}$}%
\author{Immanuel Bloch$^{1,3}$}%
\author{Christian Gross$^{1}$}%


\affiliation{$^1$Max-Planck-Institut f\"{u}r Quantenoptik, Hans-Kopfermann-Str.~1, 85748 Garching, Germany}
\affiliation{$^2$Laboratoire Charles Fabry, Institut d'optique Graduate School – CNRS – Université Paris Sud, 91127 Palaiseau, France}
\affiliation{$^3$Fakult\"{a}t f\"{u}r Physik, Ludwig-Maximilians-Universit\"{a}t M\"{u}nchen, 80799 M\"{u}nchen, Germany}%

\begin{abstract}
More than eighty years ago, H. Bethe pointed out the existence of bound states of elementary spin waves in one-dimensional quantum magnets~\cite{Bethe:1931}. To date, identifying signatures of such magnon bound states has remained a subject of intense theoretical research~\cite{caux_computation_2005, pereira_exact_2008, kohno_dynamically_2009, imambekov_one-dimensional_2012} while their detection has proved challenging for experiments. Ultracold atoms offer an ideal setting to reveal such bound states by tracking the spin dynamics after a local quantum quench~\cite{Ganahl:2012} with single-spin and single-site resolution~\cite{bakr:2010, Sherson:2010}. Here we report on the direct observation of two-magnon bound states using in-situ correlation measurements in a one-dimensional Heisenberg spin chain realized with ultracold bosonic atoms in an optical lattice. We observe the quantum walk of free and bound magnon states through time-resolved measurements of the two spin impurities. The increased effective mass of the compound magnon state results in slower spin dynamics as compared to single magnon excitations. In our measurements, we also determine the decay time of bound magnons, which is most likely limited by scattering on thermal fluctuations in the system. Our results open a new pathway for studying fundamental properties of quantum magnets and, more generally, properties of interacting impurities in quantum many-body systems.
\end{abstract}

\symbolfootnote[1]{Electronic address: {\bf takeshi.fukuhara@mpq.mpg.de}}
\maketitle%

The study of non-equilibrium processes in quantum spin models can provide fundamental insight into elementary aspects of magnetism. Magnons are the basic quasiparticle exitations around the ground state of ferromagnets and govern their low temperature physics~\cite{Wortis:1963, takahashi_one-dimensional_1971}. Due to the ferromagnetic interaction, two spin excitations can remain bound together, forming a so-called two-magnon bound state~\cite{Bethe:1931, Wortis:1963, hanus_bound_1963}. In one and two dimensions, bound states exist for all center of mass momenta, which prohibits the description of low energy properties in terms of free magnon states~\cite{Wortis:1963}. In the classical limit, magnon bound states can be regarded as the basic building blocks of magnetic solitons~\cite{fogedby_spectrum_1980, schneider_solitons_1981}.
Next to these fundamental aspects, the study of non-equilibrium dynamics in quantum spin chains is also important for a variety of applications. The evolution of two localized spin excitations realizes an interacting quantum walk~\cite{schreiber_2d_2012,lahini_quantum_2012} in the spin domain, which can be a versatile tool for the study of complex many-body systems~\cite{venegas-andraca_quantum_2012}. It is also of importance in the context of quantum information~\cite{Bose:2007}, where transport properties in a one-dimensional chain of qubits can be strongly influenced by magnon bound states~\cite{subrahmanyam_entanglement_2004}.

\begin{figure}[!t]
  \centering
  \includegraphics[width=\columnwidth]{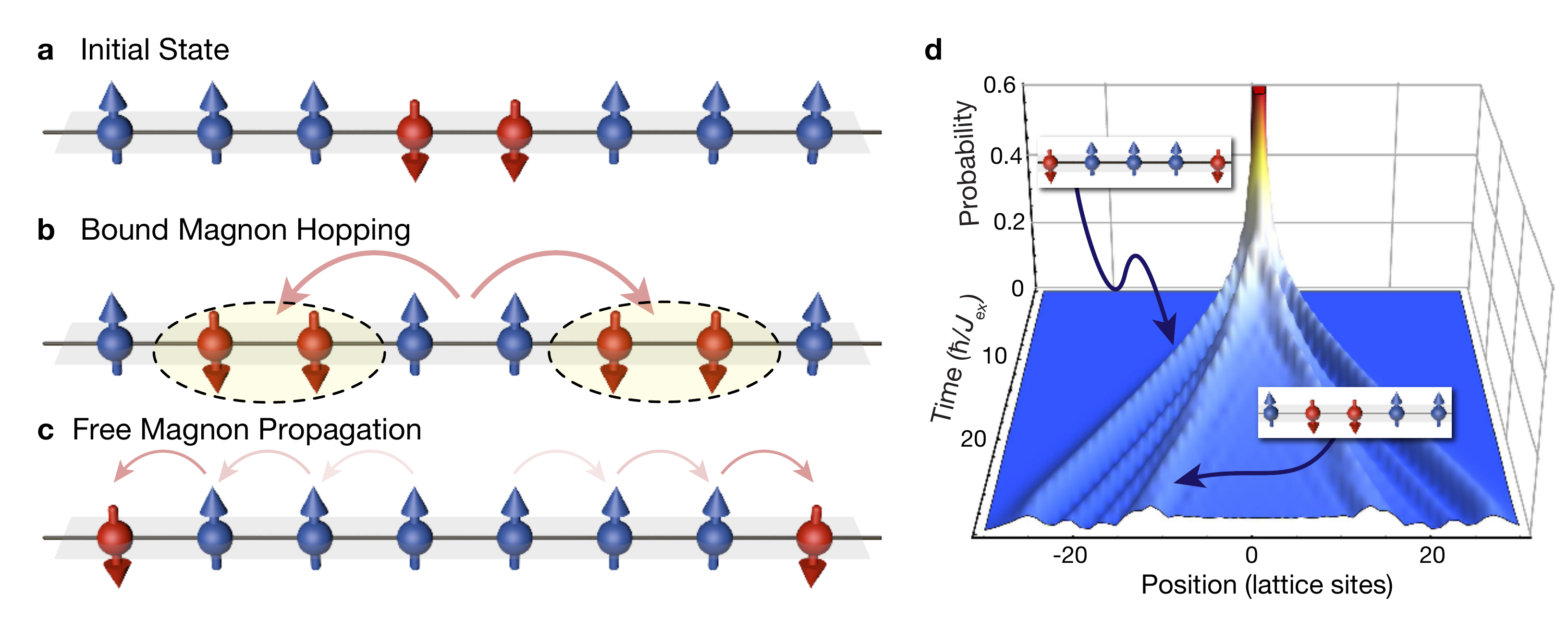}
  \caption{{\bf Schematic representation of the magnon propagation.} {\bf a-c}, Initially prepared state with two flipped spins and its decomposition into bound and free magnons propagating through the lattice. {\bf d}, Numerical results obtained from exact diagonalization showing the probability to find a flipped spin at a given lattice site following the initial state preparation. Two different wavefronts corresponding to bound and free magnons can be identified (see insets). Note that the maximum probability was clipped in the graph for clarity.}
  \label{fig:schematics}
\end{figure}

\begin{figure*}[!t]
  \centering
  \includegraphics[width=0.75\textwidth]{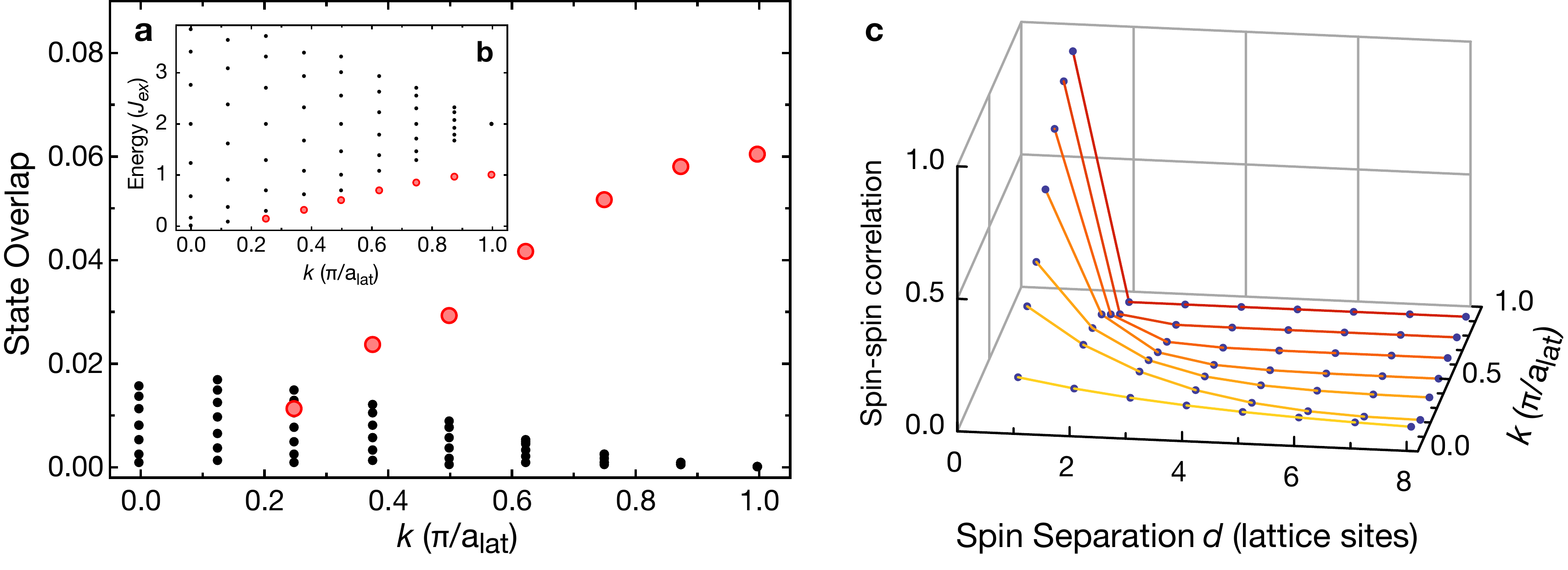}
  \caption{{\bf Quantum state analysis through Bethe Ansatz.} {\bf a}, Overlap of the initial state with the bound (red circles) and free (black dots) magnon states calculated for $N=16$ lattice sites~\cite{karbach_introduction_1998}. For illustration we show only the states with wave-vectors within the interval $k \in [0,\pi/a_{\rm lat}]$. The inset {\bf b} shows the corresponding energy spectrum. {\bf c}, Spin-spin correlation of the magnon bound states as a function of the spin separation $d=\left| j-i \right|$ for different wave-vectors $k$. For $k=\pi/a_{\rm lat}$ the wavefunction of the bound magnon state corresponds to tightly bound spins on neighbouring sites, giving the largest overlap with our initial state.}
  \label{fig:bethe}
\end{figure*}

The spin-1/2 Heisenberg model is one of the foundational models for interacting quantum spins. This model could be solved analytically in one dimension in the early 1930's by H. Bethe using a systematic Ansatz for the form of the eigenvectors~\cite{Bethe:1931}. Later, the Bethe Ansatz proved to be far more general and allowed for solving many more one-dimensional models, such as the Lieb-Liniger or the fermionic Hubbard model~\cite{batchelor_bethe_2007}, and recent, powerful extensions include the investigation of the dynamics of one-dimensional quantum many-body systems. One of the first results of Bethe's analysis was the prediction of magnon bound states.
Experimentally, infrared scattering experiments provided first evidence for the existence of such states in materials characterized by a highly anisotropic, Ising-like Hamiltonian~\cite{date_spin-cluster_1966, torrance_jr_excitation_1969}. For ultracold atoms in optical lattices, high-energy bound states have been observed in the density sector in the form of repulsively bound atom pairs~\cite{winkler_repulsively_2006, folling_direct_2007}. Optical lattice systems can also be used to realize the Heisenberg model with in principle tunable anisotropy~\cite{Kuklov:2003, Duan:2003}, where bound states occur as low energy excitations of the many-body system. Recent technological advances even allow for the in-situ control and detection of atomic spins in these experiments~\cite{weitenberg_single-spin_2011,bakr:2010,Sherson:2010}.

In our system, we make use of a one-dimensional chain of bosonic atoms in an optical lattice. Starting from an initial Mott insulating state and a fully magnetized chain, we flip two neighbouring spins in the center of the chain, thereby realizing a local quantum quench (see Fig.~\ref{fig:schematics} and refs.~\cite{Ganahl:2012,fukuhara_quantum_2013}). Making use of our single-site and single-spin resolved detection method \cite{Sherson:2010}, we are able to directly observe individual magnons and their bound states and identify both of them by correlation measurements after letting the system evolve. 

The system is described by the one-dimensional two-species single-band Bose-Hubbard Hamiltonian at unity filling. In the strong coupling limit, where the on-site interaction energy is much larger than the tunnelling matrix element, this Hamiltonian can be mapped onto the ferromagnetic spin-$1/2$ Heisenberg chain (also known as XXZ spin-$1/2$ chain) ~\cite{Kuklov:2003,Duan:2003}:
\begin{equation}
  \hat H = - J_{\rm ex} \sum_{i} \left[ \frac{1}{2} \left( \hat S_i^+ \hat S_{i+1}^- + \hat S_i^- \hat S_{i+1}^+ \right) + \Delta \hat S_i^z \hat S_{i+1}^z \right] ,
  \label{eq:Heisenberg}
\end{equation}
where $J_{\rm ex}$ is the superexchange coupling and $\Delta$ is the anisotropy between the transversal and longitudinal spin coupling. The pseudo-spin operators are defined in terms of creation $\hat a^{\dag}_ {\sigma ,i}$ and annihilation $\hat a_ {\sigma ,i}$ operators for a boson on site $i$ with spin $\sigma=\uparrow,\downarrow$: $\hat S_i^+ = \hat a^{\dag}_ {\uparrow ,i} \hat a_ {\downarrow ,i}$, $\hat S_i^- = \hat a^{\dag}_ {\downarrow ,i} \hat a_ {\uparrow ,i}$ and $\hat S_i^z = \left(\hat n_ {\uparrow ,i} - \hat n_ {\downarrow ,i} \right) / 2$, where the number operators $\hat n_ {\sigma ,i}$ count the bosons of the respective spin state on each lattice site. The transversal coupling (the first term of equation~(\ref{eq:Heisenberg})) corresponds to the spin exchange between two neighbouring sites and results in the propagation of spin excitations, or magnons~\cite{fukuhara_quantum_2013,trotzky_time-resolved_2008}. The longitudinal coupling describes a nearest-neighbour interaction between the spins, which favours ferromagnetic order for $J_{\rm ex} \Delta > 0$. This term is the origin of the magnon bound states: two flipped spins can lower their energy when located on neighbouring sites. For our scattering parameters, the Heisenberg interactions are almost isotropic, that is $\Delta \simeq 1$ (see Supplementary Information). We note that the Heisenberg model above can be mapped onto a spinless Fermi system with nearest-neighbour attractive interactions via the Jordan-Wigner transformation~\cite{Book:Giamarchi}. In the noninteracting case ($\Delta=0$) magnons therefore behave as free fermions.

Starting from a general  wavefunction for the case of two flipped spins of the form
\begin{equation}
  |\Psi\rangle = \sum_{1\leq i < j \leq N} a(i,j) |i,j\rangle ,
\end{equation}
with $|i,j\rangle = \hat S_i^- \hat S_j^- |\ldots \uparrow \uparrow \uparrow\uparrow\ldots\rangle$ and $N$ denoting the length of the chain, we use the Bethe Ansatz to obtain the eigenvalues and eigenvectors of the system (see Fig.~\ref{fig:bethe} and ref.~\cite{karbach_introduction_1998}). The bound states can be identified from the corresponding energy spectrum through their separation from the scattering states (see Fig.~\ref{fig:bethe}b). The spatial extension of the spin-spin correlations $\sum_{i} \left|a(i,i+d)\right|^{2}$ for each bound state can be calculated (see Fig.~\ref{fig:bethe}c). This analysis reveals that our initial state $|\ldots \uparrow \uparrow \downarrow\downarrow\uparrow\uparrow\dots\rangle$ has large overlap ($\sim 50\%$) with two-magnon bound states, the rest being shared among free magnon scattering states. We therefore expect both bound and free magnon dynamics to appear in the subsequent dynamical evolution after flipping two neighbouring spins (see Fig.~\ref{fig:schematics}).

\begin{figure*}[!t]
  \centering
  \includegraphics[width=0.75\textwidth]{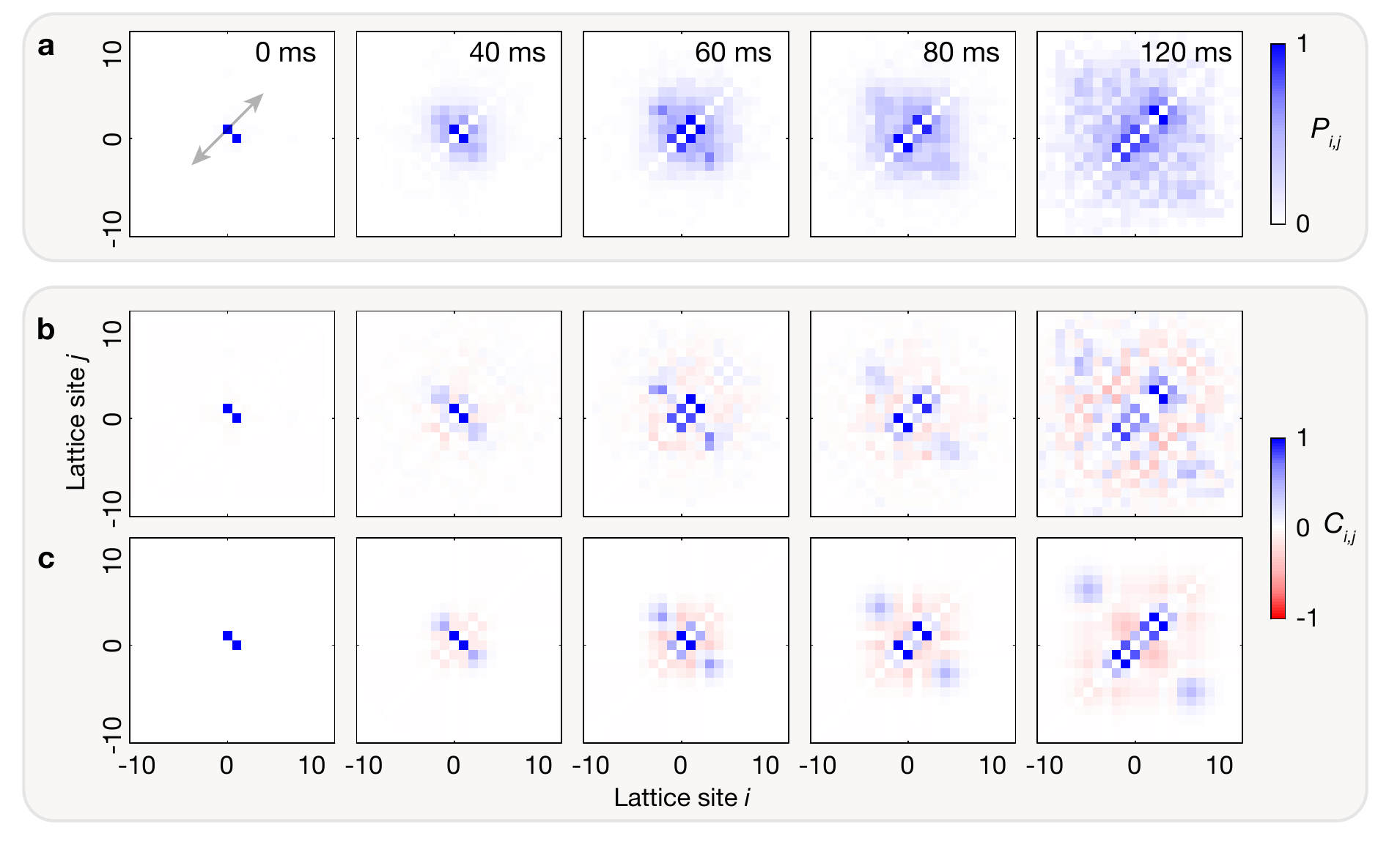}
  \caption{{\bf Spatial correlations after dynamical evolution.} {\bf a}, Measured joint probability distributions $P_{i,j}$ of the position of the two spins for different evolution times as indicated. The bound magnon signal and its spreading is visible on the diagonals $j = i \pm 1$ (arrow). Color scales are normalized for each image to the measured peak value. {\bf b}, Corresponding correlation functions $C_{i,j}=P_{i,j}-P_i P_j$ of the measured data. The subtraction of uncorrelated detection events caused by finite temperature effects and finite preparation fidelity gives better access to the correlation signal of the zero-temperature two-magnon evolution. For example, anti-bunching for the free magnons becomes visible, which is reflected in the outward propagating signal along the orthogonal diagonal. {\bf c}, Numerical results for the correlations using exact diagonalization. Color scales are normalized analog to {\bf a}. Note that the symmetry around the $j=i$ diagonal in all plots is given by construction.}
  \label{fig:correlation}
\end{figure*}

The experiment started with the preparation of a two-dimensional degenerate gas of Rubidium-87 atoms in the \up state in a single antinode of a vertical optical lattice (lattice spacing $a_{\rm lat}=532$\,nm). The spin degree of freedom was encoded in two hyperfine states with $\up\equiv\ket{F=1,m_{F}=-1}$ and $\down \equiv \ket{2,-2}$. By ramping up one horizontal lattice, the gas was then split into approximately ten decoupled one-dimensional tubes of comparable length. The splitting was carried out in $120$\,ms with a final lattice depth of $30\,E_r$, where $E_{r}=h^2/(8m a_{\rm lat}^2)$ denotes the recoil energy and $m$ is the atomic mass. Simultaneously, the lattice along the tubes was increased to $V=20\,E_r$, driving the system into the Mott insulating phase. In the next step, we applied a microwave driven spin flip to the state \down of two neighbouring atoms at the center of each chain. For this spin flip, we used a line-shaped laser beam generated with a spatial light modulator that selectively shifted the addressed sites in resonance with the microwave radiation~\cite{fukuhara_quantum_2013}. The addressing light was chosen at a wavelength and polarization such that the \up states were unaffected while the \down states were lowered in energy, and thus pinned at their positions. We then ramped down the lattice along the tubes to $V=10E_{r}$ in $50\,\mathrm{ms}$ and subsequently switched off the addressing beam within $1\,\mathrm{ms}$. This marked the starting point of the dynamics. At this final lattice depth, the dynamics is sufficiently fast ($J_{\rm ex}/\hbar= 54$\,Hz) compared to the typical heating time of several $100\,\mathrm{ms}$. After a variable evolution time, we rapidly ramped up all lattices to approximately $80\,E_r$ in order to freeze out the dynamics. For state selective detection, we applied a microwave sweep to invert the spin population followed by a resonant laser pulse on the closed cycling transition in order to push out the $\ket{2, -2}$ majority component. We finally detected the atoms originally in the \down state (now mapped to the remaining $\ket{1,-1}$ state) with single-site resolution~\cite{Sherson:2010}.

We analysed the extracted atom positions in terms of a joint probability $P_{i,j}$ to simultaneously detect atoms on lattice sites $i$ and $j$ along the tubes. Only data sets with exactly two atoms per tube were included. Approximately $50\%$ of the data are discarded through this process, mainly due to the finite spin flip fidelity. In Fig.~\ref{fig:correlation}a, we show the resulting probability distributions. The bound state population is directly reflected in the strong signal along the diagonals $j = i \pm 1$. The spread along this direction increases with evolution time, which is a signature for the correlated motion of the spin pair forming the bound state. To subtract uncorrelated detection events caused by finite temperature effects and finite preparation fidelity (see Supplementary Information), we calculate the correlation function $C_{i,j}=P_{i,j}-P_{i}P_{j}$, where $P_{i}=\sum_j P_{i,j}$ is the probability to find one atom on site $i$ (see Fig.~\ref{fig:correlation}b). Due to the hard-core constraint, trivial anti-correlations are present for $i=j$, which we discard in the analysis. Next to the strong signal of bound magnons, a second feature is visible along the orthogonal diagonal. It corresponds to those free magnon states with which the prepared initial state has finite overlap (see Fig.~\ref{fig:bethe}). As we show below, these are spins detected at largest distance from each other given by the maximal free magnon velocity $J_{\rm ex} a_{lat}/\hbar$. Their anti-bunching behaviour of propagating in opposite directions can be understood intuitively from the aforementioned mapping of the Heisenberg model to a fermionic Hamiltonian. Numerical results based on exact diagonalization of the Heisenberg chain (assuming zero temperature), shown in Fig.~\ref{fig:correlation}c, are in remarkable agreement with the experimental data. Analog to the free magnon wavefront, the one for bound magnons spreads also at its maximum velocity $J_{\rm ex} a_{lat}/(2\hbar \Delta)$ due to a singularity in the probability density of propagation velocities (see Supplementary Information and ref.~\cite{Ganahl:2012}).

\begin{figure}[t!]
  \centering
  \includegraphics[width=\columnwidth]{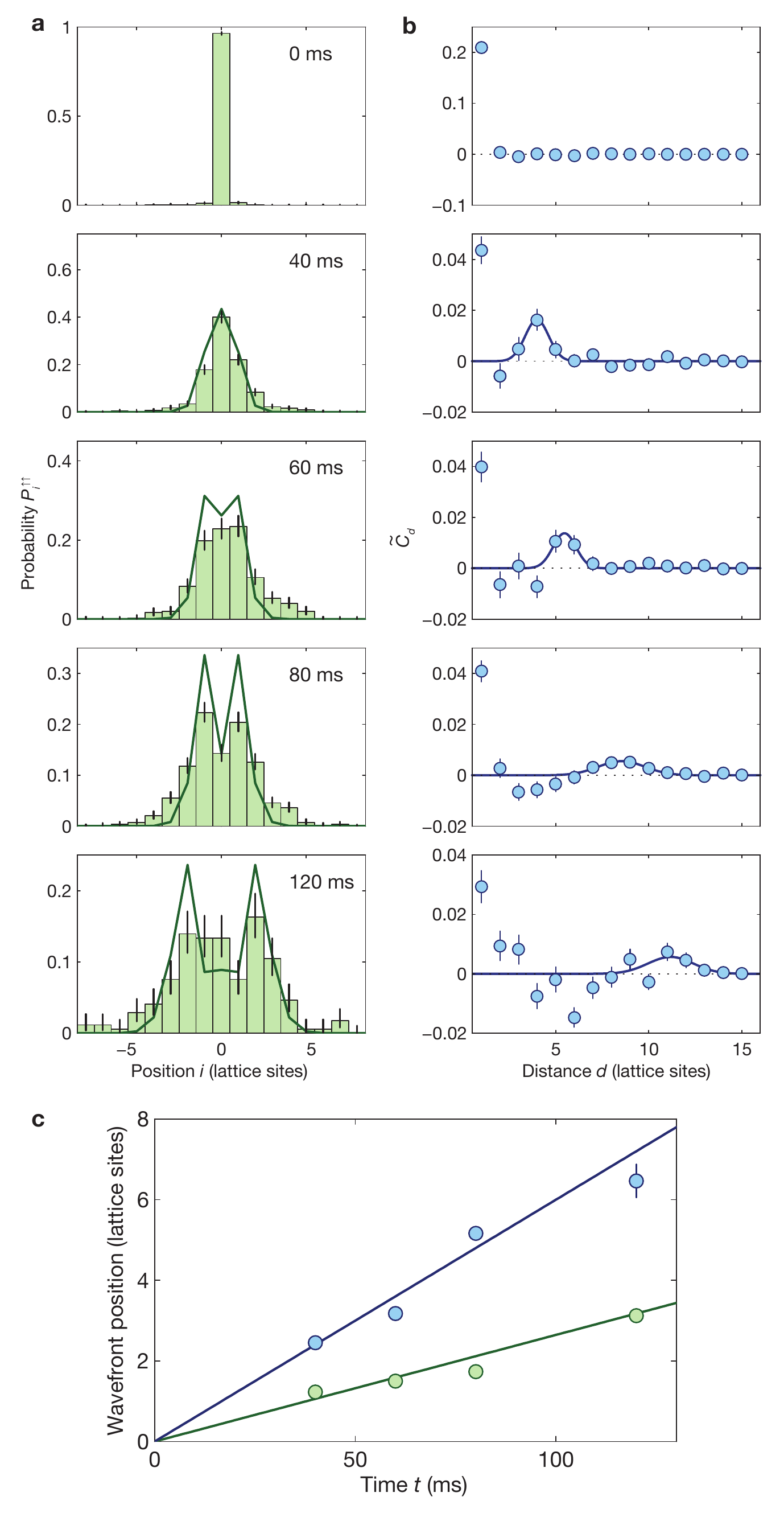}
  \caption{{\bf Spreading wavefront velocity of bound and free magnons.} {\bf a}, Bound state probability distributions $P_{i}^{\uparrow \uparrow}$ for different evolution times. The green bars show the experimental data. We extract the widths via Bessel function fits to the data (solid green lines). {\bf b}, Propagation of the free magnons. The extracted correlation functions $\widetilde{C}_{d}$ versus distance $d$ are plotted for the same evolution times as used in {\bf a} (blue circles). The signal at $d=1$ is due to the bound states while the outwards moving peak stems from free magnons. The position and width of this peak are captured by Gaussian fits (dark blue lines). {\bf c}, Comparing the propagation velocities. Linear regression of the extracted widths for the bound states (green) yields a velocity of $26(^{+2 +6}_{-2})$ sites/s compared to $60(3)$ sites/s for the wavefront of the free magnons. Error bars represent one s.e. of the mean.}
  \label{fig:propagation_velocity}
\end{figure}

 To investigate the dynamics of the magnon bound states in more detail, we concentrate on the diagonals $j=i \pm 1$ in Fig.~\ref{fig:correlation} and analyse the evolution of the normalized distribution $P_{i}^{\uparrow \uparrow}=P_{i,i+1}/\sum_j P_{j,j+1}$, that is we use only data where two atoms on adjacent sites have been detected. We expect the magnon bound states to spread as compound objects almost freely across the lattice. We therefore extract the width $w$ of the distributions $P_{i}^{\uparrow \uparrow}$ by fitting the data with Bessel functions of the first kind $\left[ {\cal J}_i \left( w \right)\right]^2$ (ref.~\cite{fukuhara_quantum_2013}). To measure the propagation velocity of the free magnon excitations, we analyse the correlations $\widetilde{C}_{d}=\sum_{i} C_{i,i+d}$ as a function of distance, shown in Fig.~\ref{fig:propagation_velocity}b. Here, the correlation signal at $d=1$ is due to the magnon bound states, while the second positive correlation signal, at a distance increasing with evolution time, is the free magnon contribution. We determine the position of the free magnons via Gaussian fits and define the wavefront as the center plus one Gaussian sigma to take the dispersion into account (see Supplementary Information). Figure~\ref{fig:propagation_velocity}c shows the measured wavefront positions of both the free and bound magnon states versus time. A linear fit yields the velocities $v_{f} = 60(3)\,\mathrm{sites/s}$ for the free and $v_{b} = 26(^{+2 +6}_{-2})\,\mathrm{sites/s}$ for the bound magnons, where the first uncertainty of $v_{b}$ is due to the fit and the second one takes a systematic underestimation of the bound state velocity into account (see Supplementary Information). The ratio of the two velocities is $v_{f}/v_{b} = 2.3(^{+0.2}_{-0.7})$, consistent with the predicted value $v_{f}/v_{b}=2\Delta=2$ for the isotropic case~\cite{Ganahl:2012}. 
 
\begin{figure}[t!]
  \centering
  \includegraphics[width=0.9\columnwidth]{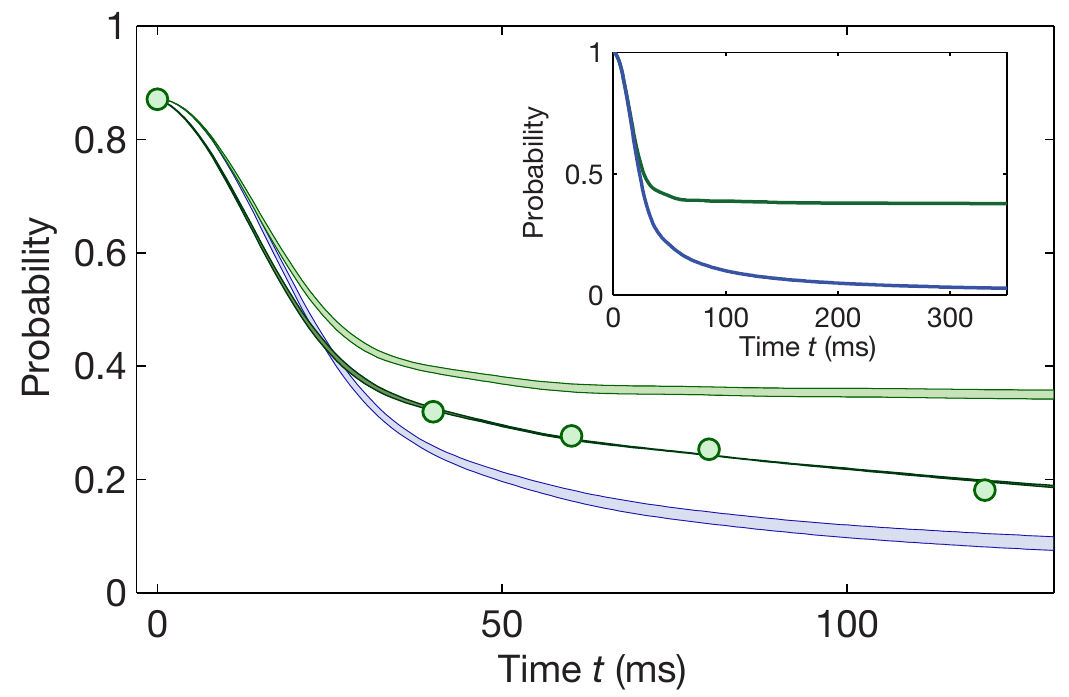}
  \caption{{\bf Stability of the bound state.} Probability to find two spins at neighbouring sites as a function of the evolution time. The green circles are the experimental data and statistical error bars are smaller than the circles. We show numerical calculations (exact diagonalization) for the isotropic $\Delta =1$ case (green shaded area) and for $\Delta =0$ (blue shaded area), taking the preparation fidelity of 87\% and the resulting uncertainty into account. The darker green line is a fit based on the isotropic numerical result multiplied by an exponential decay. Inset: Numerical prediction for longer evolution time and without correcting for the preparation fidelity. The nearest-neighbour probability approaches zero in the $\Delta=0$ case (blue line) while it converges to a finite value of 38\% for $\Delta=1$ (green line).}
  \label{fig:probability}
\end{figure} 

Above we analysed the data in the context of the isotropic Heisenberg chain and found good agreement with the theoretical predictions. However, the experiment was not carried out at zero temperature, resulting in a finite density of particle or hole excitations (approximately $10\%$) in the atomic chain. We expect that coupling of these thermal excitations to the magnon bound states leads to a finite lifetime. To extract this lifetime, we plot the probability to find two atoms on adjacent sites ($\sum_i P_{i,i+1}$) versus time in Fig.~\ref{fig:probability}. For comparison, we show the zero temperature prediction of the Heisenberg chain for the isotropic experimental case and for the case $\Delta=0$, for which no bound states exist. Here, we take the finite preparation fidelity ($87(1)\%$) for flipping the spin of two atoms at adjacent sites into account (see Supplementary Information). For long evolution times (inset), the probability approaches zero for the non-interacting case ($\Delta=0$), while it reaches a finite value of 38\% in the isotropic model. This value is smaller than the overlap between our initial state and the magnon bound states because of the finite extension of the bound states beyond neighbouring sites (see Fig.~\ref{fig:bethe}c). We find the experimental data to lie in between the two scenarios (see Fig.~\ref{fig:probability}). We fit the data with a heuristic model, which assumes the numerical prediction of the isotropic Heisenberg chain multiplied by an exponential decay. The extracted decay time of the bound magnon state is $\tau = 210(20)$\,ms, where the uncertainty includes both the fitting error and the uncertainty in the numerical prediction. We believe this decay time to be determined by both thermal density fluctuations that are present already initially and technical heating during the evolution dynamics. It remains an interesting challenge for future theory work to explain the lifetime due to the interaction of bound magnons with density fluctuations on the spin chain. 

In conclusion, we deterministically realized a local quantum quench in a Heisenberg spin chain. We microscopically tracked the resulting dynamics and directly observed distinctive magnon bound state correlations and their evolution with time. This is the first realization of an interacting quantum walk in a magnetic spin chain. From the quantum simulation perspective, our results also constitute the first observation of interacting spins in optical lattices. Future studies might address the question of the stability of magnon bound states in an environment containing thermal as well as stronger quantum fluctuations or even the binding of two impurities in a superfluid environment, where one expects a ``bi-polaron'' to form. Other interesting extensions would be the study of universal Efimov physics using three magnons~\cite{Nishida:2013}. The reported results also pave the way towards the deterministic microscopic engineering of complex magnetic many-body states and the study of magnetic correlations in non-equilibrium situations.

\section*{Acknowledgements}
We thank H. G. Evertz, M. Haque, J.-S. Caux and W. Zwerger for discussions. We thank J. Zeiher for proofreading the manuscript. This work was supported by MPG, DFG, EU (NAMEQUAM, AQUTE, Marie Curie Fellowship to M.C.), and JSPS (Postdoctoral Fellowship for Research Abroad to T.F.).


\bibliography{References}





\section*{Supplementary Information}

\renewcommand{\thefigure}{S\arabic{figure}} 
\setcounter{figure}{0}

\section{Experimental procedure}
The general experimental procedure closely followed the one published in~\cite{fukuhara_quantum_2013}. Additionally, for the long evolution time ($120$\,ms) experiments presented here, we used a vertically propagating, blue-detuned ($\sim 667$\,nm) beam to reduce the harmonic confinement in the horizontal plane. This enabled us to create larger Mott insulating plateaus with unity filling and thereby avoid reflections of the magnons from the boundaries of the atomic spin chains. The typical length of the chains was 20 sites with the deconfinement and 13 without. We generated this deconfinement beam by a broadband superluminescent diode in order to avoid possible interference of the beam with reflections from the vacuum window. The beam was successively amplified by two tapered amplifiers.

\section{Extraction of the wavefront velocity from the fits}
The validity of our method to extract the wavefront velocities was checked by analysing the results obtained from simulated data. To extract the velocity of the bound magnons, we used fits with Bessel functions. The Bessel function is not the exact distribution to describe the evolution of the bound magnons that we prepare, but, as we show below, it is suitable to capture the position of the wavefront in the distributions (see Fig.~S1a). The positions extracted from the simulated data for different times are shown in Fig.~S1b. In the long-time average, the resulting velocity agrees with the theoretical prediction ($ J_{\rm ex} a_{\rm lat}/2 \hbar$) for the bound magnons with $\Delta = 1$. For shorter, experimentally accessible times, the fits can underestimate the velocity up to $20\%$. This is the reason for the systematic error on $v_{b}$ given in the main text.

The free magnon velocity is extracted from the time evolution of the position and width of the outward moving peak in the correlation functions $\widetilde{C}_d$ by using Gaussian fits: $A \exp\left[ - \left( d - c \right)^2/ s^2 \right]$. To focus on the propagating peak, we exclude from the fit both the points at $d=1$, which show the strong positive signal of the bound state, and the points with negative values. In Fig.~S2, the peak position $c$ and the wavefront position $c+s$ are plotted. The wavefront velocity yields twice the velocity of a single free magnon since two free magnons propagate separately in opposite directions. The deviation of the extracted velocity from the maximal velocity $J_{\rm ex} a_{\rm lat}/ \hbar$ of the single free magnon is found to be only $3\%$.

\begin{figure}
  \centering
  \includegraphics[width=\columnwidth]{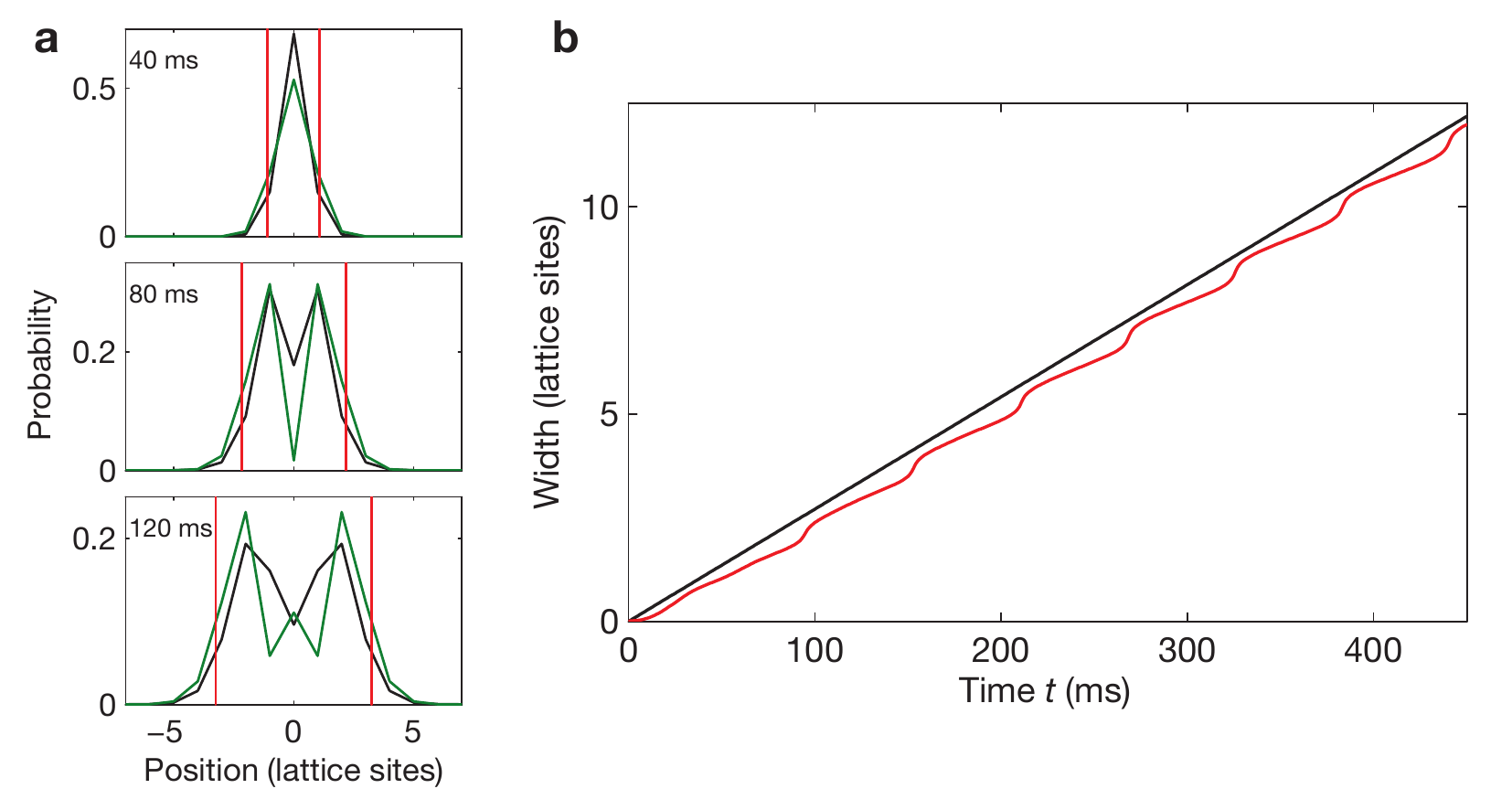}
  \caption{{\bf Propagation of bound magnons.} {\bf a} Calculated probability distribution $P_{i}^{\uparrow \uparrow}$ (black lines) together with the Bessel function fit (green lines) for different evolution times ($40$, $80$, and $120$\,ms). The red vertical lines show the width extracted from the fit. {\bf b} Determination of the velocity. The red line is the extracted width from the Bessel function fit versus the evolution time. The black line corresponds to the expected maximum velocity $\left( J_{\rm ex} a_{\rm lat}/2 \hbar \right)$.}
  \label{fig:bound_magnon_si}
\end{figure}

\begin{figure}
  \centering
  \includegraphics[width=0.8\columnwidth]{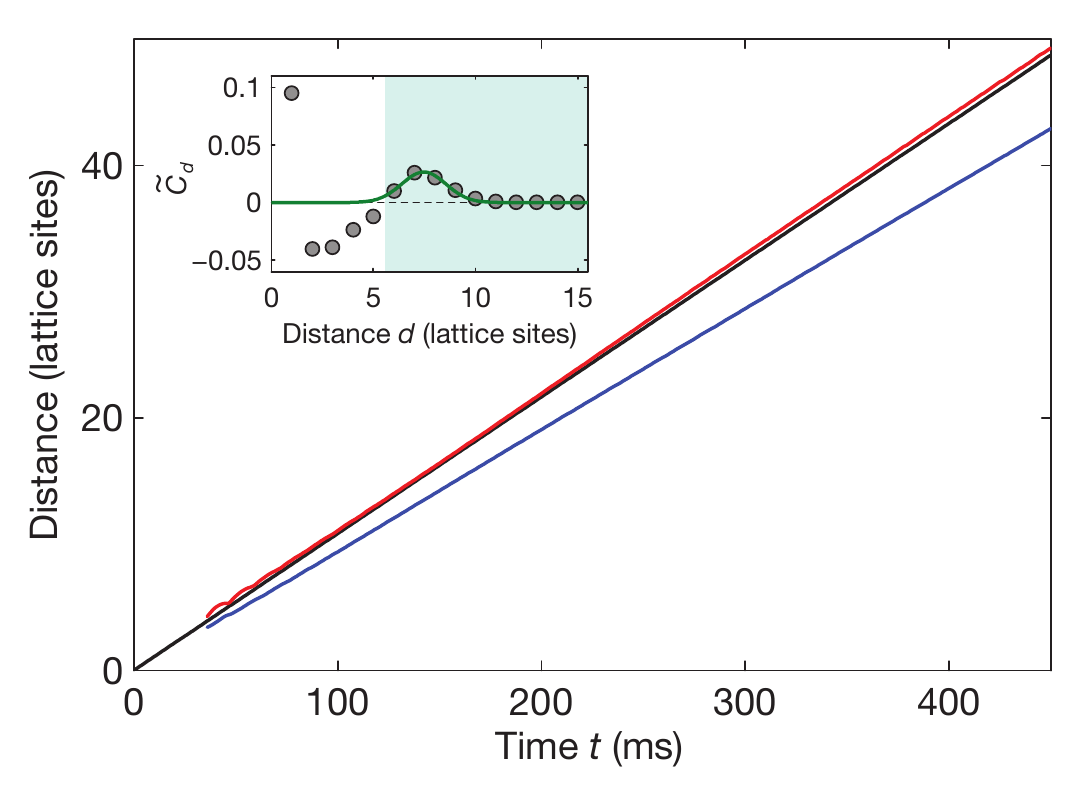}
  \caption{{\bf Propagation of free magnons.} The blue and red lines show the Gaussian center $c$ and the center plus the width $c+s$. Note that the center moves slower than the maximum wavefront velocity. The black line, almost overlapping with the red line, corresponds to twice the expected single magnon maximum velocity $\left(2 J_{\rm ex} a_{\rm lat}/ \hbar \right)$. The inset shows an example of the Gaussian fit (green line). The gray circles represent the calculated correlation function $\widetilde{C}_d$ for the evolution time of $80$\,ms. The blue shade highlights the region used for the fit.}
  \label{fig:unbound_magnon_si}
\end{figure}

\section{Preparation fidelity}
The preparation fidelity for flipping the spin of two atoms on neighbouring sites is estimated to be 87\%. This value is limited by two factors. First, the spin-flipping process might have addressed two spins initially separated by a larger distance. Second, the flipping process might have succeeded for one atom only, while the second atom observed is one from the majority component that was not removed during the push-out process because of its finite efficiency (98–99\%). These two effects have different contributions on the probability to find two atoms on neighbouring sites after the evolution time. In the first case, the effect can be calculated by solving the dynamics with the measured initial distributions ($P_{i,j} \left( t=0 \right)$). For the second case, we can assume that falsely measured atoms are uniformly distributed over the chain (they were generated after the dynamics). The calculated probabilities shown in Fig.~\ref{fig:probability} take both these effects into account. The width of the shaded region displayed in Fig.~\ref{fig:probability} is due to the uncertainty of the ratio between the two effects.

\section{Parameters of the Heisenberg model}
The superexchange coupling $J_{\rm ex}$ and the anisotropy $\Delta$ are given by~\cite{Kuklov:2003,Duan:2003,GarciaRipoll:2003,Altman:2003a}
\begin{align}
J_{\rm ex} &= \frac{4 J_{\uparrow} J_{\downarrow}}{U_{\uparrow \downarrow}}, \\
J_{\rm ex} \Delta &= \left( \frac{4 J_{\uparrow}^2}{U_{\uparrow \uparrow}} + \frac{4 J_{\downarrow}^2}{U_{\downarrow \downarrow}} -2 \frac{J_{\uparrow}^2 + J_{\downarrow}^2}{U_{\uparrow \downarrow}} \right).
\end{align}
Here $J_{\sigma}$ are tunnelling matrix elements for a boson with spin $\sigma=\uparrow,\downarrow$ and $U_{\sigma \sigma'}$ are the on-site interaction energies between bosons with spin $\sigma$ and $\sigma'$. In our case, the tunnelling matrix elements are spin independent ($J_{\uparrow} = J_{\downarrow} = J$), and the interaction energies are almost the same ($U_{\uparrow \uparrow} \approx U_{\downarrow \downarrow} \approx U_{\uparrow \downarrow} = U$). The anisotropy is $\Delta = 0.986$ for our ratios of the interaction energies ($U_{\uparrow \uparrow} : U_{\downarrow \downarrow} : U_{\uparrow \downarrow} = 100.4:99.0:99.0$) that follow from the respective scattering lengths~\cite{Pertot:2010,Hoefer:2011}.

\section{Numerical calculations using exact diagonalization}
We calculated the dynamics of the effective Heisenberg chain by directly diagonalizing the Hamiltonian. Since the number of magnons, or the total magnetization, is conserved, we considered only the Hilbert space containing two magnons. The numerical calculation was done for a superexchange coupling of $J_{\rm ex}/\hbar= 54$\,Hz, which has been estimated from $J_{\rm ex}=4 J^{2}/U$. Here, the tunnelling matrix element $J$ was obtained from the observation of the quantum walk of a single free atom, as shown in \cite{weitenberg_single-spin_2011}. The on-site interaction energy $U$ was obtained from an ab initio band-structure calculation using lattice depths, that were calibrated from amplitude modulation spectroscopy. For the simulations we used open boundary conditions and lattice sizes of 61 or 81 sites, depending on the evolution time, making sure that the magnons remain sufficiently far away from the edges to avoid spurious reflections.

\section{Density of states and initial distribution of group velocities}
Our initially prepared state $\ket{\Psi_i}$ can be decomposed into two parts:
\begin{equation}
\ket{\Psi_i}=\ket{\Psi_b}+\ket{\Psi_f}\nonumber,
\end{equation}
that describe the overlap with magnon bound states ($\ket{\Psi_b}$) and free magnon scattering states ($\ket{\Psi_f}$). The bound magnon part is expanded in bound magnon eigenstates $\ket{\psi_k}$ with center-of-mass wave-vector $k$ as
\begin{equation}
\ket{\Psi_b} \propto \int_k dk \langle \psi_k|\Psi_b \rangle\ket{\psi_k}.
\end{equation}
The probability density $P(v)$ to find our initial state in magnon bound states that have a group velocity $v$ can be written as:
\begin{align}
P(v)&\propto\bra{\Psi_b}\int_k dk\, \delta(v-v_g(k))\ket{\psi_k}\bra{\psi_k}\ket{\Psi_b}\nonumber\\
&=\int_k dk\, \delta(v-v_g(k))|\langle \psi_k |\Psi_b \rangle|^2 \nonumber\\
&=\frac{1}{\sqrt{v_{max}^2-v^2}}\sum_{k_v} |\langle \psi_{k_v}|\Psi_b \rangle|^2.
\end{align}
The group velocity is $v_g(k)=\frac{d\epsilon}{dk}=\frac{J}{2}\sin(k)$ with the bound state dispersion $\epsilon=\frac{J}{2}(1-\cos(k))$~\cite{Wortis:1963, karbach_introduction_1998}. The maximum group velocity is $v_{max}=J/2$. The sum in the last line runs over all wave-vectors $k_v$ that yield a particular group velocity $v_g(k_v)=v$. 
The quantity $ |\langle \psi_{k}|\Psi_b \rangle|^2$ describes the probability to find the initial state in a magnon bound state with $k$ and is plotted in Fig.~\ref{fig:bethe}a. It is nonzero for the values $k=\pm\pi/2$ that yield the maximum group velocity $v_g(k)=\pm v_{max}$. Therefore, $P(v)$ shows a divergence for $v=\pm v_{max}$. 

The quantity $ \int_k \!dk\,\, \delta(k-v_g(k)) \propto \frac{1}{\sqrt{v_{max}^2-v^2}}$ essentially describes the density of states for a particular group velocity $v$ \cite{Ganahl:2012}. The singularity in this density of states is the origin of the singularity in $P(v)$.
 
\end{document}